\documentclass[conference]{IEEEtran}
\IEEEoverridecommandlockouts

\usepackage{cite}
\usepackage{amsmath,amssymb,amsfonts}
\usepackage{algorithmic}
\usepackage{graphicx}
\usepackage{textcomp}
\usepackage{xcolor}
\usepackage{url}

\def\BibTeX{{\rm B\kern-.05em{\sc i\kern-.025em b}\kern-.08em
    T\kern-.1667em\lower.7ex\hbox{E}\kern-.125emX}}
\begin{document}

\title{Incorporating Talker Identity Aids With Improving Speech Recognition in Adversarial Environments\\
}

\author{\IEEEauthorblockN{Sagarika Alavilli*, Annesya Banerjee*, Gasser Elbanna*, Annika Magaro*}
\IEEEauthorblockA{
\textit{Speech and Hearing Bioscience and Technology} \\
\textit{Harvard University}\\
\textit{MA, USA} \\
\texttt{\{salavill, annesyab, gelbanna, amagaro\}@mit.edu}
\thanks{$^{\ast}$ Equal contribution. Names are sorted alphabetically based on Surnames.}
}
}
\maketitle

\begin{abstract}
Current state-of-the-art speech recognition models are trained to map acoustic signals into sub-lexical units. While these models demonstrate superior performance, they remain vulnerable to out-of-distribution conditions such as background noise and speech augmentations. In this work, we hypothesize that incorporating speaker representations during speech recognition can enhance model robustness to noise. We developed a transformer-based model that jointly performs speech recognition and speaker identification. Our model utilizes speech embeddings from Whisper and speaker embeddings from ECAPA-TDNN, which are processed jointly to perform both tasks. We show that the joint model performs comparably to Whisper under clean conditions. Notably, the joint model outperforms Whisper in high-noise environments, such as with 8-speaker babble background noise. Furthermore, our joint model excels in handling highly augmented speech, including sine-wave and noise-vocoded speech. Overall, these results suggest that integrating voice representations with speech recognition can lead to more robust models under adversarial conditions.
\end{abstract}

\begin{IEEEkeywords}
voice identification, automatic speech recognition, joint training
\end{IEEEkeywords}


\section{Introduction}

Linguistic and acoustic information are essential elements of human speech. Although it is recognized that voice and talker characteristics influence speech perception~\cite{mullennix1989some}, research has traditionally examined these aspects separately. This divide is also evident in automatic speech recognition (ASR) systems, which prioritize linguistic content and often overlook speaker-specific acoustic cues. However, emerging evidence shows that familiarity with talker-specific features can significantly improve speech comprehension and recognition~\cite{nygaard1994speech, nygaard1998talker, yonan2000effects, holmes2024does}.

Deep learning-based ASR models are typically trained to convert acoustic signals into lexical (words)~\cite{whisper} or sub-lexical (characters or phonemes)~\cite{baevski2020wav2vec} outputs. While they perform well in controlled settings, such as clear audio or reading speech~\cite{yang2021superb}, their accuracy drops significantly with minimally altered inputs, like sine-wave or noise-vocoded speech, or in noisy conditions~\cite{shah2024speech}. We hypothesize that these vulnerabilities arise partly because the representations learned by these models are incentivized to be invariant to indexical (non-speech) information.

In this study, we evaluate the performance of Whisper~\cite{whisper}, a state-of-the-art speech recognition model, under adversarial conditions by introducing background speaker babble and applying sine-wave and noise-vocoding transformations. We find that Whisper's performance deteriorates significantly even with slight adversarial manipulations. To address this, we propose a model trained on a dual task of speech and speaker recognition, encouraging the retention of both speech and non-speech information in its representations. By combining speech embeddings from the pre-trained Whisper model with speaker embeddings from the ECAPA-TDNN model~\cite{desplanques2020ecapa}, our ``joint" model demonstrates improved robustness. Our experiments reveal the relevance of speaker representations for generalizable speech recognition.

\begin{figure*}[htbp]
\centering
\includegraphics[width=\textwidth,height=0.4\textheight]{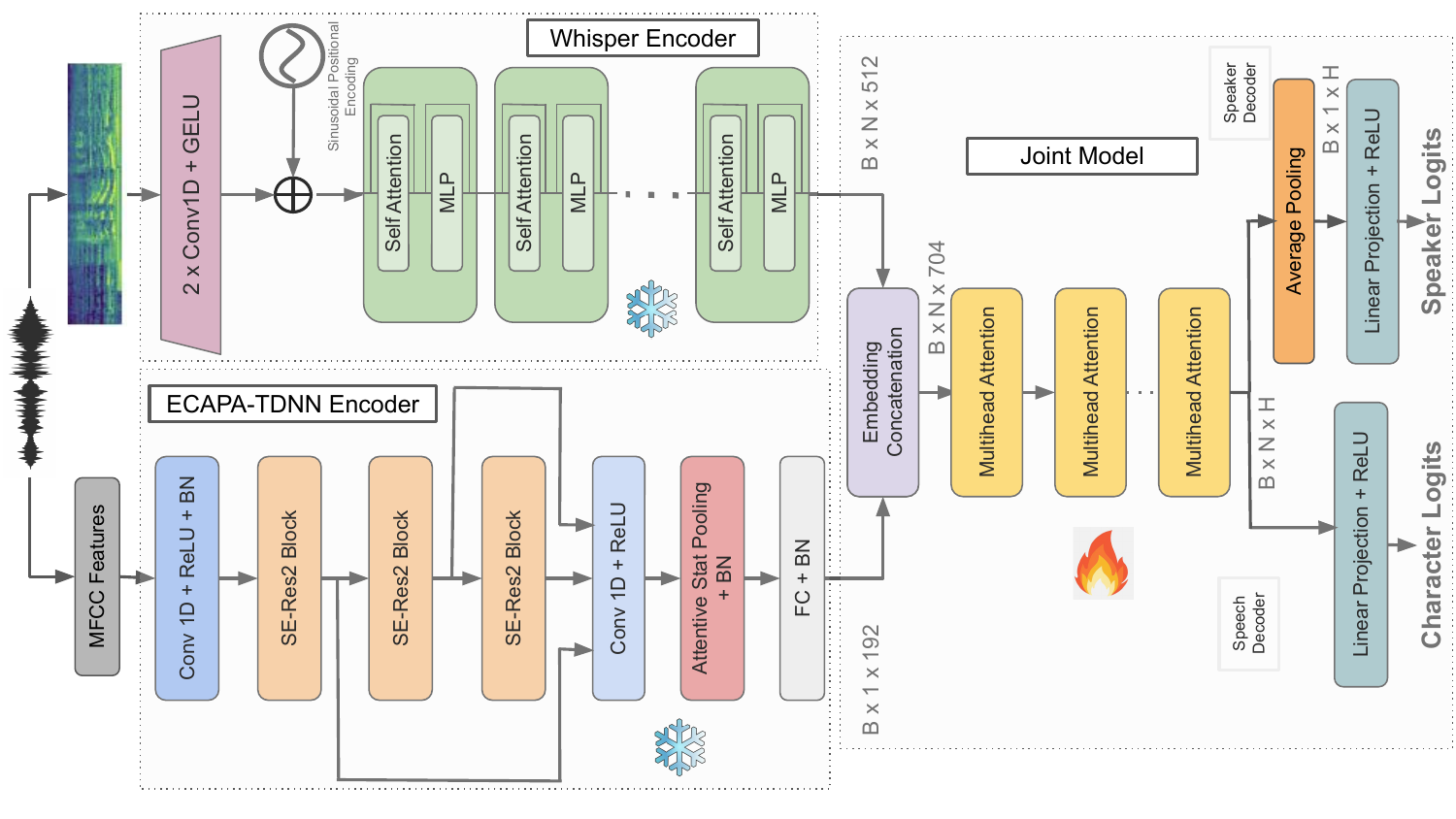}  
\caption{Architecture of dual-task Joint model. The linear layer, ReLU activation, and mapping layer (to corresponding task logits), as described in the text, are represented together as the Linear Projection + ReLU stage. B, N, and H denote the batch size, the number of time frames, and the number of feed-forward nodes in the transformer layer, respectively.}
\label{fig}
\end{figure*}

\section{Methods}
\subsection{Candidate Models}
\noindent \textbf{Whisper} \cite{whisper} is a supervised, transformer-based encoder-decoder model that takes mel spectrograms as an input with 30 sec long audio samples. This work used the \texttt{huggingface} pre-trained model \texttt{openai/whisper-base} with $74M$ parameters and was trained on $680K$ hours of transcribed audio. The penultimate layer is used to extract embeddings $\mathbf{e}_w \in \mathrm{R}^{512}$.\\\\

\noindent \textbf{ECAPA-TDNN} is a state-of-the-art model for speaker recognition \cite{desplanques2020ecapa}, pre-trained on $7205$ speakers from the VoxCeleb 1 and 2 corpora \cite{nagrani2020voxceleb, chung2018voxceleb2}. It processes MFCCs ($\in \mathrm{R}^{80}$) from audio windows of $25ms$ with a hop size of $10ms$. The model uses stacked Squeeze-Excitation Res2Blocks, followed by a feature aggregation layer, and applies attentive statistical pooling. The pooled features are then fed into a linear layer and an AAM-softmax output layer, with the penultimate layer being $\mathbf{e}_{ec} \in \mathrm{R}^{192}$ speaker embeddings.\\\\

\noindent \textbf{Joint Model.} We developed a multi-layer transformer model combining ECAPA-TDNN and Whisper embeddings. The concatenated 704-dimensional embeddings are processed through multi-head transformers, producing embeddings of size $H$, where $H$ is the number of feed-forward nodes. For speech recognition, these embeddings are passed through a linear layer with ReLU activation, mapped to class logits, and optimized with a CTC loss \cite{graves2006connectionist}. For speaker recognition, the transformer outputs are average-pooled, passed through a linear layer with ReLU activation, and mapped to talker logits with cross-entropy loss. Both losses are summed for training, with Whisper and ECAPA-TDNN weights kept frozen. \\\\

\noindent \textbf{Whisper (Character).} Unlike the original Whisper model that predicts the next word, our joint model was trained to predict the next character due to data limitations. We also trained a Whisper (Character) model to mitigate this difference and predict the next character. The approach was precisely the same as the joint model, except that the Whisper (Character) model did not have access to any speaker embedding. In this model, only the original Whisper embeddings of dimension 512 were passed through the stack of transformers and then to the speech decoder layers.

\begin{figure*}
\centering
\includegraphics[width=\textwidth]{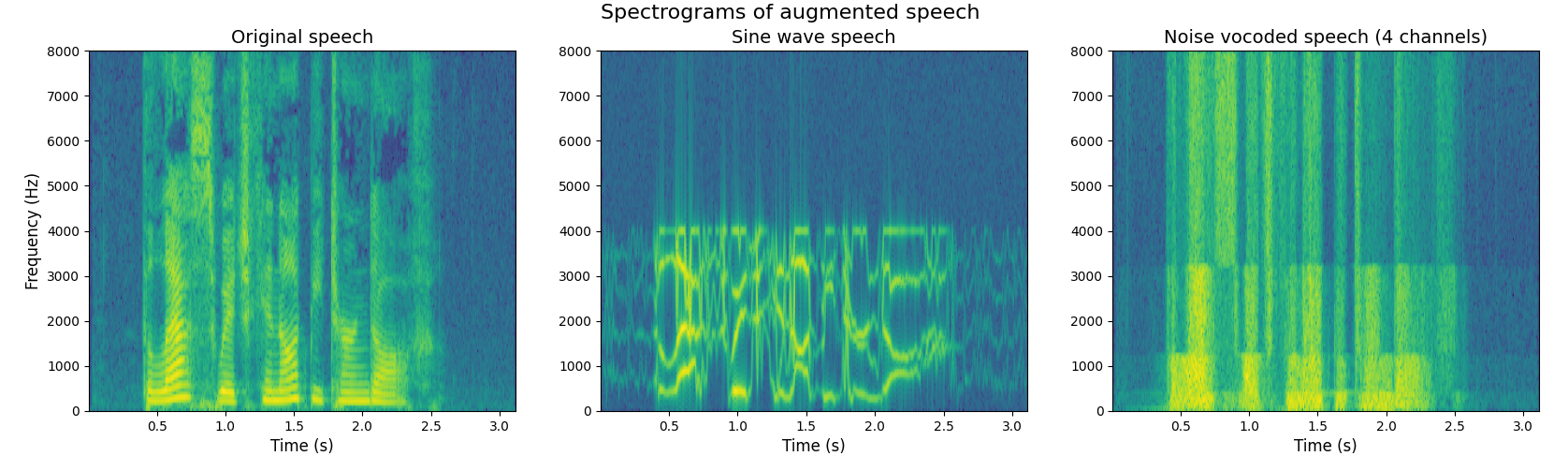}  
\caption{Example spectrograms of an original speech excerpt, sine wave speech, and noise-vocoded speech. (Left) Original speech excerpt. (Middle) Sine wave speech, generated with four bands from the original speech excerpt. (Right) noise-vocoded speech, generated with four channels from the original speech excerpt.}
\label{fig}
\end{figure*}

\subsection{Experiments}
\noindent \textbf{8 Speaker Babble Experiment.}
\label{babble_methods}
It has previously been established that speech recognition models are vulnerable to background noise \cite{shah2024speech}. However, it is possible that a model trained to represent voice along with speech might generalize better to testing in noise. By training a model on speech and speaker identification tasks, our model might be better at identifying a target voice among background noise.

To test this, we tested both Whisper models and our joint model on a series of speech clips embedded in 8-speaker babble. We varied the signal-to-noise ratio (SNR) from -15 to 20 dB, in steps of 5 dB, for 1000 background-foreground pairings. Every model was presented with all background-foreground pairs at every SNR. We calculated the character error rate (CER) as our model outputs a character (Equation \ref{cer}).

\begin{equation}
CER = \frac{substitutions + deletions + insertions}{n\ characters} \label{cer}
\end{equation}

\noindent \textbf{Augmented speech experiment.}
\label{augmented_methods}
Humans are robust to different forms of augmented speech. Two well-known speech augmentations are sine wave speech and noise-vocoded speech. We tested the joint model and Whisper's performance on these two augmentations. Sine wave speech was generated by dividing a speech signal into four formants. The center frequencies of these formants were then found and were replaced with sine-wave modulated center frequencies (Fig. 2). To create noise-vocoded speech, a speech signal was filtered into several subbands or channels, and the amplitude envelopes of these channels were applied to noise. Then, the subbands were added together, resulting in a noise-vocoded signal. This type of speech processing simulates the processing that a cochlear implant does  \cite{shannon1995speech}. The noise-vocoded speech was generated with 1, 4, 16, and 64 channels (Fig. 2). We calculated the CER for Whisper and the joint model for each condition.

\subsection{Datasets}
\noindent \textbf{Training Data.} For training the joint model, we used the Common Voice (English) dataset  \cite{ardila2019common}. The texts were lowercase, and punctuation was removed. The texts were further tokenized before passing on to the model. Following data preprocessing, the vocabulary size was 30 (English letters 'a' to 'z,' space, padding, the start and end of the sentence). We used utterances from 200 speakers, i.e., a 200-class classification task for speaker recognition. The $100$ male and $100$ female speakers with the highest frequency of occurrence in the dataset were selected. For speakers with more than $400$ occurrences in the dataset, we randomly sampled $400$ occurrences. This resulted in a set of speakers with between $300$ and $400$ occurrences in the dataset. This same dataset was used to train the Whisper (Character) model. For evaluation, we used stimuli from the test split of the Common Voice dataset.\\\\

\noindent \textbf{8 Speaker Babble Data.} To test the impact of background noise on the models, we created a dataset of 1000 speech-babble pairs. Both the speech and 8-talker speech babble were randomly sampled from the test split of the Common Voice dataset. For each speech-babble pair, we combined them at dB SNRs from -15 to 20, increasing in 5 dB increments. We also tested models on clean speech at infinite SNR. All speech-babble pairs were generated at all SNRs to ensure that the acoustic difficulty of the signal only varied from SNR. \\

\begin{figure}
\includegraphics[width=0.5\textwidth]{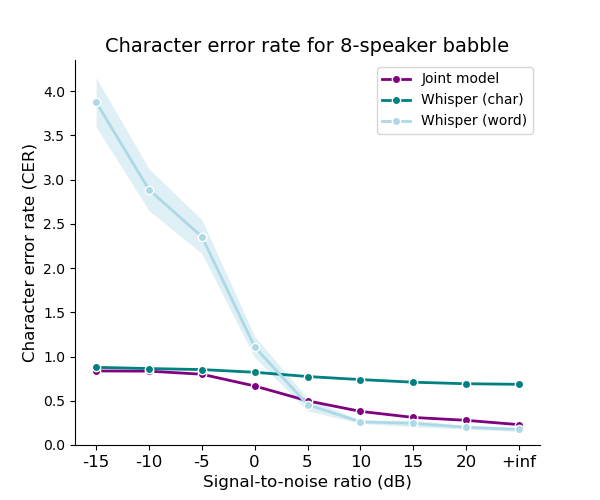}  
\caption{Comparison of model speech recognition performance speech in 8-speaker babble. Character error rate is plotted from SNRs ranging from -15 dB to 20 dB in increments of 5, with the +inf label representing speech without noise. }
\label{fig}
\end{figure}
\noindent \textbf{Augmented speech.} Speech augmentations were applied to a subset of the general evaluation dataset. For noise-vocoded speech, 1000 speech excerpts were randomly sampled from the evaluation dataset. These were transformed into noise-vocoded speech with 1, 4, 16, and 64 channels. For sine wave speech, 795 speech excerpts were randomly sampled from the evaluation dataset and were transformed into sine wave speech with four bands.\\

\section{Results}
\subsection{Speech and Speaker Recognition Performance} \label{task_results}
We first trained and evaluated our joint model performance on the speech and speaker recognition tasks. To do so, we trained multiple variants of the joint model, each with a slightly different architecture. Table \ref{tab:saganet} shows the joint model performance for our top three example architectures. Here, we varied the number of multi-head transformers used in the stack and the number of feed-forward nodes in each transformer layer. Note that we used 8 heads for each transformer in all cases. The joint model speaker recognition performance was consistently high across all the variants. The word error rate was also low but showed notable differences across architecture variants. Therefore, we chose the architecture variant 2 as our best model. This model's WER is $18.30$\%, comparable to the original Whisper (Base) model's WER on the Common Voice dataset. Joint model variant 2 was used for all further analyses reported here. Additionally, we used the variant 2 architecture specifications to train our Whisper (Character) model. 

\begin{table}[htbp]
\caption{Joint Model Performance: Word Error Rate (WER) and Speaker Recognition Accuracy (SRA)}
\begin{center}
\begin{tabular}{|c|c|c|c|c|}
\hline
\textbf{Architecture} & \textbf{Transformer} & \textbf{Feed-forward} & \textbf{WER} $\downarrow$ & \textbf{SRA} $\uparrow$ \\
 & \textbf{Layers} & \textbf{Nodes} &  &  \\
\hline
Variant 1 & 2 & 1024 & 29.10\% & 99.40\% \\
\hline
Variant 2 & 4 & 512 & \textbf{18.30\%} & 99.50\% \\
\hline
Variant 3 & 4 & 1024 & 21.30\% & \textbf{99.60\%} \\
\hline
\multicolumn{4}{l}{$^{\mathrm{*}}$Bold values indicate the best performance.}
\end{tabular}
\label{tab:saganet}
\end{center}
\end{table}

\subsection{8 Speaker Babble Experiment} \label{babble_results}
To analyze the impact of background noise on speech recognition, we plotted CER as a function of SNR for Whisper and our joint model (Fig. 3). We see that Whisper outperforms our model at infinite SNR. Both models perform comparably well at high SNRs. However, we see that Whisper begins dropping in performance at around 5 dB SNR much more rapidly than our model. At -10 dB SNR, our model has an average CER of 1, while Whisper's is at 3. This suggests that learning a joint representation of voice and speech can result in a model that is more robust to background noise.

\subsection{Augmented Speech Experiment}
\label{augmented_results}
We analyzed the joint model's speech recognition performance vs. Whisper on two forms of augmented speech: noise-vocoded speech and sine-wave speech. Both of these types of augmentations make the voice unrecognizable, but humans are generally more robust than models. We hypothesized that the joint model's more human-like representation would yield greater robustness to these conditions. 

We first plotted the CER of the joint model and Whisper as a function of the number of channels in the noise-vocoded speech. We found that Whisper performed slightly better than the joint model for the 16-channel and 64-channel conditions of noise-vocoded speech. However, for the 1-channel and 4-channel noise-vocoded conditions, augmentations that deviate much further from normal speech, the joint model had much higher performance than Whisper, with Whisper's CER reaching almost 2. The joint model's CER stays below 1. Similarly, we see better generalization to sine-wave speech in the joint model (Fig. 4).

\begin{figure}
\includegraphics[width=0.5\textwidth]{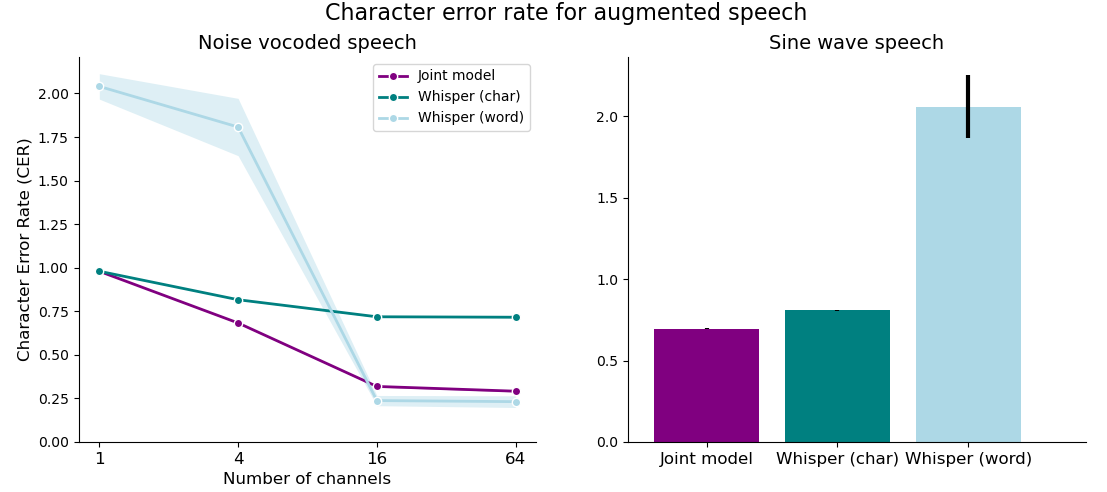}  
\caption{Comparison of model speech recognition performance on two forms of augmented speech. (Left) Performance on noise-vocoded speech. (Right) Performance on sine wave speech. }
\label{fig}
\end{figure}

\section{Discussion and Conclusion}

In this work, we hypothesized that integrating speaker representations could improve speech recognition under adversarial conditions. We developed a transformer-based model combining speaker and speech embeddings to perform speech and speaker recognition. The joint model was tested against the pre-trained Whisper model, and our model's variant focused only on speech recognition. As shown in Figure 3, the joint model matches Whisper's performance at high signal-to-noise ratios (SNRs) but outperforms it at lower SNRs. This improvement is likely due to Whisper's focus on word prediction, which struggles with noisy inputs. To ensure a fair comparison, we also tested a variant of our model that generates characters without speaker recognition.

Results showed that while the character-generating Whisper variant performed better than the original Whisper at low SNRs, both versions were still outperformed by the joint model trained on both speaker and speech recognition tasks. These findings suggest that incorporating speaker identity representations can be beneficial for speech recognition, particularly in noisy conditions. The joint model might leverage voice representations to maintain speaker tracking, which may facilitate more accurate speech recognition even in adverse acoustic environments.

Furthermore, we evaluated the models using various types of augmented speech to simulate degraded signal conditions. As shown in Figure 4, a similar performance pattern emerged, consistently demonstrating the superiority of the joint model when handling degraded signals. This further reinforces the potential advantage of incorporating speaker identity representations for robust speech recognition in challenging acoustic environments.

Our joint model, trained on both speaker and speech recognition tasks, outperformed Whisper in handling noisy and out-of-distribution speech. This work highlights the value of incorporating speaker information into speech recognition models. Rather than aiming for speaker invariance, leveraging speaker-specific representations can enhance robustness under challenging conditions like noise and signal degradation. Integrating speaker identity cues allows models to adapt better to acoustic variations, improving recognition accuracy and reliability. This approach offers new possibilities for developing more resilient speech recognition systems.

\section{Acknowledgments}
We thank Lakshmi Govindarajan and Ajani Stewart for their comments and feedback on the manuscript.

\bibliographystyle{IEEEtran}
\bibliography{ref}

\begin{thebibliography}{10}
\providecommand{\url}[1]{#1}
\csname url@samestyle\endcsname
\providecommand{\newblock}{\relax}
\providecommand{\bibinfo}[2]{#2}
\providecommand{\BIBentrySTDinterwordspacing}{\spaceskip=0pt\relax}
\providecommand{\BIBentryALTinterwordstretchfactor}{4}
\providecommand{\BIBentryALTinterwordspacing}{\spaceskip=\fontdimen2\font plus
\BIBentryALTinterwordstretchfactor\fontdimen3\font minus \fontdimen4\font\relax}
\providecommand{\BIBforeignlanguage}[2]{{%
\expandafter\ifx\csname l@#1\endcsname\relax
\typeout{** WARNING: IEEEtran.bst: No hyphenation pattern has been}%
\typeout{** loaded for the language `#1'. Using the pattern for}%
\typeout{** the default language instead.}%
\else
\language=\csname l@#1\endcsname
\fi
#2}}
\providecommand{\BIBdecl}{\relax}
\BIBdecl

\bibitem{mullennix1989some}
J.~W. Mullennix, D.~B. Pisoni, and C.~S. Martin, ``Some effects of talker variability on spoken word recognition,'' \emph{The Journal of the acoustical society of America}, vol.~85, no.~1, pp. 365--378, 1989.

\bibitem{nygaard1994speech}
L.~C. Nygaard, M.~S. Sommers, and D.~B. Pisoni, ``Speech perception as a talker-contingent process,'' \emph{Psychological Science}, vol.~5, no.~1, pp. 42--46, 1994.

\bibitem{nygaard1998talker}
L.~C. Nygaard and D.~B. Pisoni, ``Talker-specific learning in speech perception,'' \emph{Perception \& psychophysics}, vol.~60, no.~3, pp. 355--376, 1998.

\bibitem{yonan2000effects}
C.~A. Yonan and M.~S. Sommers, ``The effects of talker familiarity on spoken word identification in younger and older listeners.'' \emph{Psychology and aging}, vol.~15, no.~1, p.~88, 2000.

\bibitem{holmes2024does}
E.~Holmes, ``How does voice familiarity affect speech intelligibility?'' \emph{The Journal of the Acoustical Society of America}, vol. 155, no. 3\_Supplement, pp. A263--A263, 2024.

\bibitem{whisper}
A.~Radford, J.~W. Kim, T.~Xu, G.~Brockman, C.~McLeavey, and I.~Sutskever, ``Robust speech recognition via large-scale weak supervision,'' in \emph{International conference on machine learning}.\hskip 1em plus 0.5em minus 0.4em\relax PMLR, 2023, pp. 28\,492--28\,518.

\bibitem{baevski2020wav2vec}
A.~Baevski, Y.~Zhou, A.~Mohamed, and M.~Auli, ``wav2vec 2.0: A framework for self-supervised learning of speech representations,'' \emph{Advances in neural information processing systems}, vol.~33, pp. 12\,449--12\,460, 2020.

\bibitem{yang2021superb}
S.-w. Yang, P.-H. Chi, Y.-S. Chuang, C.-I.~J. Lai, K.~Lakhotia, Y.~Y. Lin, A.~T. Liu, J.~Shi, X.~Chang, G.-T. Lin \emph{et~al.}, ``Superb: Speech processing universal performance benchmark,'' \emph{arXiv preprint arXiv:2105.01051}, 2021.

\bibitem{shah2024speech}
M.~A. Shah, D.~S. Noguero, M.~A. Heikkila, and N.~Kourtellis, ``Speech robust bench: A robustness benchmark for speech recognition,'' \emph{arXiv preprint arXiv:2403.07937}, 2024.

\bibitem{desplanques2020ecapa}
B.~Desplanques, J.~Thienpondt, and K.~Demuynck, ``Ecapa-tdnn: Emphasized channel attention, propagation and aggregation in tdnn based speaker verification,'' \emph{arXiv preprint arXiv:2005.07143}, 2020.

\bibitem{nagrani2020voxceleb}
A.~Nagrani, J.~S. Chung, W.~Xie, and A.~Zisserman, ``Voxceleb: Large-scale speaker verification in the wild,'' \emph{Computer Speech \& Language}, vol.~60, p. 101027, 2020.

\bibitem{chung2018voxceleb2}
J.~S. Chung, A.~Nagrani, and A.~Zisserman, ``Voxceleb2: Deep speaker recognition,'' \emph{arXiv preprint arXiv:1806.05622}, 2018.

\bibitem{graves2006connectionist}
A.~Graves, S.~Fern{\'a}ndez, F.~Gomez, and J.~Schmidhuber, ``Connectionist temporal classification: labelling unsegmented sequence data with recurrent neural networks,'' in \emph{Proceedings of the 23rd international conference on Machine learning}, 2006, pp. 369--376.

\bibitem{shannon1995speech}
R.~V. Shannon, F.-G. Zeng, V.~Kamath, J.~Wygonski, and M.~Ekelid, ``Speech recognition with primarily temporal cues,'' \emph{Science}, vol. 270, no. 5234, pp. 303--304, 1995.

\bibitem{ardila2019common}
R.~Ardila, M.~Branson, K.~Davis, M.~Henretty, M.~Kohler, J.~Meyer, R.~Morais, L.~Saunders, F.~M. Tyers, and G.~Weber, ``Common voice: A massively-multilingual speech corpus,'' \emph{arXiv preprint arXiv:1912.06670}, 2019.

\end{thebibliography}

\end{document}